\documentstyle[seceq,epsf,wrapft]{ptptex}

\DeclareRobustCommand\openone{\leavevmode\hbox{\small1\normalsize\kern-.33em1}}

\title{Response functions of gapped spin systems in high magnetic field}

\author{
Alexei K. {\sc Kolezhuk}$^{1,2}$
and 
Hans-J\"urgen {\sc Mikeska}$^{2}$}

\inst{
$^{1}$ Institute of Magnetism, National Academy of Sciences and Ministry of
Education,
36(B) Vernadskii avenue, 03142 Kiev, Ukraine\\ 
$^{2}$ Institut f\"ur Theoretische Physik, Universit\"at Hannover,\\
Appelstra{\ss}e 2, 30167 Hannover, Germany}

\recdate{
\today
}

\abst{We study the dynamical structure factor of  gapped
one-dimensional spin systems
in the critical phase in high magnetic field. It is shown that the
presence of a ``condensate'' in the ground state in the high-field
phase leads to interesting signatures in the response functions. 
}

\begin{document}

\maketitle

\section{Introduction}

Recently, there has been an increasing interest in the physics of 
gapped one-dimensional spin systems subject to an external magnetic
field which is strong enough to close the gap and bring the system
into a new critical phase with finite magnetization and gapless
excitations.
\cite{SakaiTakahashi91,Chaboussant+98a,Chaboussant+98,Honda+98,Honda+99,Orendac+99,Totsuka98,Mila98,ChitraGiamarchi97,GiamarchiTsvelik99,FurusakiZhang99,HikiharaFurusaki01,Bocquet+01} 
The low-energy physics of the critical phase is
usually described by the Luttinger liquid,
\cite{ChitraGiamarchi97,KonikFendley01} in some cases the problem can
be directly mapped to the effective $S={1\over2}$ chain.
\cite{Totsuka98,Mila98,GiamarchiTsvelik99,KMS99} Such a description is
equivalent to neglecting certain degrees of freedom (e.g., two of the
three rung-triplets in case of the strongly coupled spin
ladder\cite{Totsuka98,Mila98,GiamarchiTsvelik99}). Those neglected states,
however, form excitation branches which contribute to the response
functions at higher energies, and this contribution is generally much
easier to see experimentally than the highly dispersed low-energy
continuum of the particle-hole (``spinon'') excitations coming from
the Luttinger liquid itself. At present, at least in two
quasi-one-dimensional materials, $\rm
Ni(C_{2}H_{8}N_{2})_{2}Ni(CN)_{4}$ (known as NENC)\cite{Orendac+99}
and $\rm Ni (C_{2}H_{8}N_{2})_{2}NO_{2}(ClO_{4})$ (abbreviated
NDMAP)\cite{Honda+99,Honda+unpub} those high-energy branches were
found to exhibit interesting behavior in electron spin resonance (ESR)
experiments, with changes in the slope of the ESR lines as functions
of the field.  Quite recently, inelastic neutron scattering (INS)
experiments on NDMAP,\cite{Zheludev+02} and on $\rm
TlCuCl_{3}$,\cite{Ruegg+02} featuring a similar behavior, were
reported.

Motivated by those observations, we have studied the ESR response in
the high-field phase of a large-D $S=1$ chain.\cite{KM02} It was shown
that presence of the Fermi sea leads to renormalization of the
energies of higher-lying excitations, which may result in nontrivial
dependence of the excitation energy on the magnetic field.

In the present paper we extend our previous work for the general case
of arbitrary transferred momentum $q$ (the ESR case corresponds to
$q=0$) and study the contributions of the high-energy excitations to
the dynamical structure factor $S(q,\omega)$ in the hardcore boson
approximation. We use the $S=1$ Affleck-Kennedy-Lieb-Tasaki
\cite{AKLT} (AKLT) chain and strongly coupled $S=1/2$ ladder as
specific examples, but the general idea behind the calculation is in
fact quite universal and applies to any gapped one-dimensional system
in the high-field critical phase. We show that characteristic
signatures in the response persist for nonzero $q$ as well, and can
therefore be observed in INS experiments.

\section{Model examples and hardcore boson approximation}

In Ref.\ \cite{KM02} we have used the hardcore boson approximation as
an effective model to study the response of the large-$D$ $S=1$
chain. In this section, we show that the same type of approximation
can be used essentially for any gapped 1D system.

\subsection{Strongly coupled ladder}

A  two-leg $S={1\over2}$ ladder with the rung coupling $J_{R}$ much
larger than the leg coupling $J_{L}$ in a strong magnetic
field $H$ can be mapped to an effective $S={1\over2}$ $XXZ$ chain by keeping
only two states on each rung, a singlet $|s\rangle$ and a triplet
$|t_{+}\rangle$, and identifying them with the effective
spin-$1\over2$ states.
\cite{Totsuka98,Mila98,GiamarchiTsvelik99} Alternatively, one may describe the
triplets $|t_{+}\rangle$ as hardcore bosons.
 The resulting $S={1\over2}$ chain model 
\begin{equation} 
\label{s12-ladder} 
\widehat{\cal H}_{S=1/2}=\sum_{n} J_{L} \Big\{
\widetilde{S}^{x}_{n}\widetilde{S}^{x}_{n+1}
+\widetilde{S}^{y}_{n}\widetilde{S}^{y}_{n+1}
+ \Delta\widetilde{S}^{z}_{n}\widetilde{S}^{z}_{n+1}
\Big\} +(J_{R}-{1\over2}J_{L}-H)\widetilde{S}^{z}_{n}
\end{equation}
has
the anisotropy $\Delta={1/2}$ of the easy-plane type and is in turn effectively
described by the Luttinger liquid model.
 If the field is only slightly
larger than the critical field $H_{c}$, the density of triplets is low
and the Luttinger parameter $K$ tends to its non-interacting value
$1$. Thus, in the vicinity of $H_{c}$ the interaction between the
bosons is irrelevant, and only the hardcore constraint is essential.

Now, if one wants to include the neglected $|t_{-}\rangle$ and
$|t_{0}\rangle$ states, one can consider them as additional species of
hardcore bosons.
Neglecting all interactions
between the bosons with the exception of the hardcore constraint, one
arrives at the effective model of the type
\begin{equation} 
\label{ha} 
\widehat{\cal H}_{\rm eff}=\sum_{n\mu} \varepsilon_{\mu}
b^{\dag}_{n,\mu}b^{\vphantom{\dag}}_{n,\mu}
+t(b^{\dag}_{n,\mu}b^{\vphantom{\dag}}_{n+1,\mu} +\mbox{h.c.}),
\end{equation}
where $\mu=+1$, $0$ and $-1$ numbers three boson species (triplet
components with $S^{z}=\mu$), $t=J_{L}$ is the hopping amplitude which
is equal for all types of particles,
$\varepsilon_{\mu}=\varepsilon_{\mu}^{(0}-\mu H$, and the zero-field
energies are given by $\varepsilon_{\mu}^{(0)}=J_{R}$.
This model is, of course,
highly simplified, but nevertheless it captures the main physics which
is hidden in the hardcore constraint.

The connection between physical spin operators and bosonic operators
for the ladder is given by the formulas
$S^{\alpha}_{n\,(1,2)}=\pm (1/2)(t_{\alpha}^{\dag} 
+t_{\alpha}^{\vphantom{\dag}})
-i\varepsilon_{\alpha\beta\gamma}t^{\dag}_{\beta}t_{\gamma}^{\vphantom{\dag}}$,
where $(\alpha,\beta,\gamma)\in(x,y,z)$ and the operators $t$ are
connected to $b$ in a standard way, $b_{\pm1}=\mp 2^{-1/2}(t_{x}\pm
it_{y})$, $b_{0}=t_{z}$.

\subsection{AKLT chain}

Mapping of the $S=1$ chain to the effective hardcore boson model is
somewhat more complicated, but still possible.
Consider $S=1$ AKLT chain in external magnetic field $H$ described by the
Hamiltonian 
\begin{equation} 
\label{AKLT} 
\widehat{\cal H}=\sum_{n}\Big\{ {\mathbf S}_{n}{\mathbf S}_{n+1} +
{1\over3}({\mathbf S}_{n}{\mathbf S}_{n+1})^{2} -H S^{z}_{n}\Big\}\,,
\end{equation}
where ${\mathbf S}_{n}$ denotes the spin-1 operator at the $n$-th
site. The zero-field gap of the AKLT model is known to be
$\Delta\simeq 0.70$ \cite{FathSolyom93}, and we are interested in the
high-field regime $H>H_{c}\equiv\Delta$, when the gap closes. 

The ground state for $H<H_{c}$ is exactly known and can be compactly
represented in the matrix product (MP) form\cite{Fannes+89,Klumper+91-93} 
\begin{equation} 
\label{gs-aklt}
 \Psi_{0}=\mbox{tr}\{g_{1}g_{2}\ldots g_{N}\}, \quad
 g_{n}=(1/\sqrt{3})(\sigma^{+}|-\rangle_{n}+\sigma^{-}|+\rangle_{n}
-\sigma^{0}|0\rangle_{n}), \nonumber
\end{equation}
where $\sigma^{\mu}$ are the Pauli matrices in the spherical basis and
$|\mu\rangle_{n}$ are the spin-1 states at the $n$-th site,
$\mu=0,\pm1$.

Excitations of the AKLT model are well approximated by
the triplet solitons (domain walls in the hidden string
order),\cite{FathSolyom93} which can be also cast
into the form of a matrix product.\cite{TotsukaSuzuki95} A soliton
which sits at the $n$-th site and has $S^{z}=\mu$ is well approximated
by the MP state
\begin{equation} 
\label{crack} 
|\mu,n)=  \mbox{tr}\{g_{1}g_{2}\ldots g_{n-1}(g_{n}\sigma^{\mu}) 
g_{n+1}\ldots g_{N}\}
\end{equation}
Soliton states
$|\mu,n)$ with different $n$ are not orthogonal.
However, one can introduce the equivalent set of states
\begin{equation} 
\label{sol}
|\mu,n\rangle =2^{-3/2}\big\{ 3|\mu,n-1)+|\mu,n) \big\}
\end{equation}
which have the orthogonality property $\langle \mu,n|
\mu',n'\rangle=\delta_{\mu\mu'}\delta_{nn'}$. The state (\ref{sol}) can be
 represented in the same MP form (\ref{crack}) with
$(g_{n}\sigma^{\mu})$ replaced by the matrices $f^{\mu}_{n}$,
\begin{eqnarray} 
\label{Msol} 
f^{+}&=&\sqrt{2\over3} \openone|+\rangle
-{1\over\sqrt{6}}(\sigma^{+}|0\rangle -\sigma^{0}|+\rangle),
\quad
f^{0}=\sqrt{2\over3} \openone|0\rangle
-{1\over\sqrt{6}}(\sigma^{+}|-\rangle
-\sigma^{-1}|+\rangle),\nonumber\\
f^{-}&=&\sqrt{2/3} \openone|-\rangle
+{1/\sqrt{6}}(\sigma^{-}|0\rangle -\sigma^{0}|-\rangle),
\end{eqnarray}
where the site index $n$ has been omitted for the sake of clarity.

One can extend this construction to many-particle states by replacing
more than one of $g$'s in (\ref{gs-aklt}) by $f$'s, however one cannot
of course get the true full Fock space in this way, because we have
four possible matrices $(g,f^{\mu})$ for every site, while the
original $S=1$ problem has only three states per site. 
The price one
has to pay is deviations from orthonormality for states with two or
more particles close to each other.
For example, one may observe that
states $|+,n;\;+,n'\rangle$, describing two solitons with $S^{z}=+1$,
remain orthogonal, but their norm depends on the distance between the
solitons,
$\langle+,n;\;+,n'|+,n;\;+,n'\rangle=1+(5/12)(-1/3)^{|n-n'|}$.
The deviations vanish  with increasing the distance between the
solitons. Therefore the approximate description in terms of
multisoliton states becomes worse as the density of solitons
increases, which means it is useful only in the
vicinity of the critical field.

Restricting our Hilbert space to the states of the MP form
(\ref{gs-aklt}) with some number of $g$ matrices replaced with $f^{+}$
ones, one can
identify the presence of the  AKLT matrix $g$ and the $S^{z}=+1$
soliton matrix $f^{+}$ at the
certain site with the effective spin-down and spin-up states, respectively.
The resulting effective $S={1\over2}$ chain is described by the
following Hamiltonian
\begin{equation} 
\label{s12-aklt} 
\widehat{\cal H}_{S=1/2}=\sum_{n} J_{\rm eff} \Big\{
\widetilde{S}^{x}_{n}\widetilde{S}^{x}_{n+1}
+\widetilde{S}^{y}_{n}\widetilde{S}^{y}_{n+1}\Big\}
+ \sum_{n}\sum_{m>0}V_{m}\widetilde{S}^{z}_{n}\widetilde{S}^{z}_{n+m}
+\sum_{n}(\varepsilon_{0}+\sum_{m}V_{m} -H)\widetilde{S}^{z}_{n}\,
\end{equation}
where $J_{\rm eff}=10/9$, $\varepsilon_{0}=50/27$, and $V_{m}$ are
very small, $V_{1}=-0.017$, $V_{2}=-0.047$, $V_{3}=0.013$,
$V_{4}=-0.0046$, etc.\cite{Krohn} Thus, if one neglects the small interaction
$V_{m}$, then in the vicinity of $H_{c}$ the AKLT chain
is effectively described by the $XY$ model, i.e. by  noninteracting
hardcore bosons.

Again, as in the case of the ladder, one can include the neglected
degrees of freedom in the hardcore approximation, arriving at the
model of the form (\ref{ha}) with
$\varepsilon_{\mu}^{(0)}=\varepsilon_{0}$ and $t=J_{\rm eff}/2$.

The connection between the physical spin operators and the bosonic
operators for the AKLT chain is more complicated than in case of
ladder. The reason for that is the fact that the actions of $S^{\mu}$
on the AKLT state are most naturally expressed in terms of the states
$|\mu,n)$ which are in turn non-locally expressed through the
orthogonalized states $|\mu,n\rangle$.  For example, the action of a
physical spin operator $S^{\mu}_{n}$ on the vacuum state $|v\rangle$
(with no bosons) is given by
\begin{equation} 
\label{sb-aklt} 
S^{\mu}_{n}|v\rangle=(2c_{\mu}/3)\big(b_{n,\mu}^{\dag}+4\sum_{m=1}^{\infty}
(-1/3)^{m})b^{\dag}_{n+m,\mu}\big)|v\rangle,
\end{equation}
where $c_{0}={1\over\sqrt{2}}$ and $c_{\pm1}=\mp1$. 

\section{Dynamical structure factor in the hardcore approximation}

Thus, we are led to the problem of calculating the response functions
of the hardcore model (\ref{ha}). The dynamical structure factor (DSF)
$S^{\alpha\alpha}(q,\omega)$ in the spectral representation has the
form
\begin{equation} 
\label{DSF} 
S^{\alpha\alpha}(q,\omega)={1\over Z}\sum_{fi}e^{-E_{f}/T}|\langle
f| S^{\alpha}(q)| i\rangle|^{2} \delta(\omega-E_{f}+E_{i}),\quad
Z=\sum_{n}e^{-E_{n}/T},
\end{equation}
where $S^{\alpha}(q)$ is the Fourier-transformed total spin
operator. The DSF is related to the dynamical susceptibility
$\chi(q,\omega)$ through the fluctuation-dissipation theorem,
\begin{equation} 
\label{fdt} 
S^{\alpha\alpha}(q,\omega)={1\over\pi}{1\over 1-e^{-\omega/T}} \Im
\chi^{\alpha\alpha}(q,\omega). 
\end{equation}
The  spin operator is generally a sum of linear and bilinear terms in
bosonic operators, which correspond to different
physical processes. 
The ground state at $H>H_{c}$ contains a ``condensate'' (Fermi sea) of $b_{+1}$
bosons, and at low temperature the most important excitations are
of the particle-hole type.  
We will thus take into account only those processes
which involve states with at most one $b_{0,-1}$ particle:
\begin{itemize}
\item[(A)]  creation/annihilation  of a low-energy $b_{+1}$ boson,
\item[(B)] 
creation/annihilation of one high-energy ($b_{-1}$ or $b_{0}$) particle, and 
\item[(C)]
transformation of a $b_{+1}$ particle into $b_{0}$ one.  
\end{itemize}

The processes of the type (A) can be considered completely within the
model of effective $S={1\over2}$ $XY$ chain, in this case there is no
need to take into account the high-energy states. 
For example, for the transversal DSF
$S^{\perp}=S^{xx}+S^{yy}$,
for $q=\pi+k$ close
to the antiferromagnetic wave number $\pi$ one can use the known
results \cite{Bocquet+01,Schulz+83-86,Barzykin01} for the dynamical
susceptibility $\chi^{xx}(q,\omega)=\chi^{yy}(q,\omega)$ which yields
\begin{equation} 
\label{spin12} 
\chi^{xx}(\pi+k,\omega)=A_{x}(H){\Gamma^{2}(3/4) v^{1/2}\over 4(\pi
T)^{3/2}}  
{\Gamma\left({1\over8} -i{\omega-vk\over4\pi T} \right)
\Gamma\left({1\over8} -i{\omega+vk\over4\pi T} \right)
\over
\Gamma\left({7\over8} -i{\omega-vk\over4\pi T} \right)
\Gamma\left({7\over8} -i{\omega+vk\over4\pi T} \right)
}\,.
\end{equation}
Here $A_{x}(H)$ is the non-universal amplitude which is known
numerically \cite{HikiharaFurusaki01}, and $v$ is the Fermi velocity. 
A similar expression is
available for the longitudinal susceptibility;\cite{Bocquet+01} for
the longitudinal DSF of the XY chain in case of zero temperature a
closed exact expression is available as well \cite{Mueller+81}, and
for $T\not=0$ the exact longitudinal DSF can be calculated
numerically.\cite{Derzhko+00} Applying (\ref{fdt}), one obtains in
this way the contribution $I^{A}(q,\omega)$ of the (A) processes into
the dynamic structure factor.  This contribution describes a
low-energy ``spinon'' continuum.

The processes of (B) and (C) type, however, cannot be analyzed in the
language of $S={1\over2}$ chain and require going back to the hardcore
boson problem.

\subsection{(B)-type transitions}

Let us consider first the zero temperature case to
understand the main features.

For the case when not more than one high-energy particle is
present, the model (\ref{ha}) can be solved
exactly. \cite{CastellaZotos93} One deals essentially with a mobile
``impurity'' in the hardcore boson system; creation of the impurity
leads to the orthogonality catastrophe \cite{Anderson67} and to the
corresponding Fermi-edge type singularity in the response.

In absence of the impurity, the eigenstates have a determinantal form
\begin{equation} 
\label{eig} 
|\Psi_{k_{1}\cdots k_{N}}\rangle
=\sum_{x_{1}x_{2}\cdots x_{N}}(1/\sqrt{N!})
 |\det\{ \psi_{i}(x_{j})\}|,
\end{equation}
where $\psi_{i}(x)={1\over\sqrt{L}} e^{ik_{i}x}$ are the free plane
wave functions, $N$ is the number of $b_{+1}$ particles (let us assume
it to be even, for definiteness), $x_{j}$ denote their positions, and
$L$ is the system length. The absolute value sign above (and
throughout the paper) should be
understood as a shorthand for having the antisymmetric sign factor
attached to the determinant, which ensures symmetry of the wave
function under permutations.  The allowed values of momenta $k_{i}$
are given by
\begin{equation} 
\label{allow} 
k_{i}=\pi+(2\pi/L)I_{i},\quad i=1,\ldots,N
\end{equation}
where the numbers $I_{i}$ should be all different and half-integer
(integer if $N$ is odd). The energy of this state is
$E=\sum_{i=1}^{N}(\varepsilon_{+1}+2t\cos k_{i})$, and the total
momentum  $P=\sum_{i}k_{i}$ is zero ($\mbox{mod} 2\pi$) when $N$ is
even. The ground state is defined by the Fermi sea configuration with
$I_{i}$ running from $-(N-1)/2$ to $(N-1)/2$, which corresponds to the
momenta in the interval $[k_{F},2\pi-k_{F}]$ with the Fermi momentum
defined as
\begin{equation} 
\label{kF} 
k_{F}=\pi(1-N/L)
\end{equation}

The excited configuration with one ``impurity'' boson $b_{\mu}$ having
the momentum $\lambda$
can be also represented in the determinantal form \cite{CastellaZotos93}
\begin{equation} 
\label{exc} 
 |(\mu,\lambda)_{k_{1}'\cdots k_{N}'}\rangle
={1\over\sqrt{L}}\sum_{x_{0}} e^{iP'x_{0}}
|x_{0}\rangle_{k_{1}'\cdots k_{N}'},\quad
 |x_{0}\rangle_{k_{1}'\cdots k_{N}'}=\sum_{y_{1}\cdots y_{N}} 
{1\over\sqrt{N!}} |\det\{ \varphi_{i}(y_{j}) \}|\,.
\end{equation}
Here $x_{0}$ denotes the coordinate of the impurity, 
and $y_{j}=x_{j}-x_{0}$ are
the coordinates of other particles relative to the impurity. 
the functions $\varphi_{j}(y)$ are given by\cite{CastellaZotos93}
\[
\varphi_{j}(y)=A_{j}[e^{i(k'_{j}y+\delta_{j})}
-(\sin\delta_{j}/\sum_{l}\sin\delta_{l})
\sum_{l}e^{i(k'_{l}y+\delta_{l})}],
\]
where the phase shifts $\delta_{j}$ are in our case of noninteracting
hardcore particles independent of $k'_{j}$ and are all equal to
$-\pi/2$, and $A_{j}$ are the normalization factors.
The functions $\varphi_{j}(y)$
were shown\cite{CastellaZotos93} 
to be in the thermodynamic limit $L\to\infty$ asymptotically equivalent to
the free scattering states
$\varphi_{j}(y)\simeq {1\over\sqrt{L}} e^{i(k'_{j}y+\delta_{j})}$.
The energy $E'$ and the total momentum  $P'$ of the excited state are given by
\begin{equation} 
\label{ep-exc} 
E=\sum_{j=1}^{N} (\varepsilon_{+1}+2t\cos k'_{j})
+\varepsilon_{\mu}+2t\cos\lambda,\qquad 
P'=\sum_{j=1}^{N} k'_{j}+\lambda\,.
\end{equation}
The allowed wave vectors $k'_{j}$ and $\lambda$ are given by the same
formula (\ref{allow}), but since the total number of particles has
changed by one, they are different from those of the
ground state: if $N$ is even, the numbers $I_{j}$ in (\ref{allow}) are
half-integer for the ground state, but integer for the excited state
(\ref{exc}). 

\begin{figure}
\centerline{\epsfxsize=60mm\epsfbox{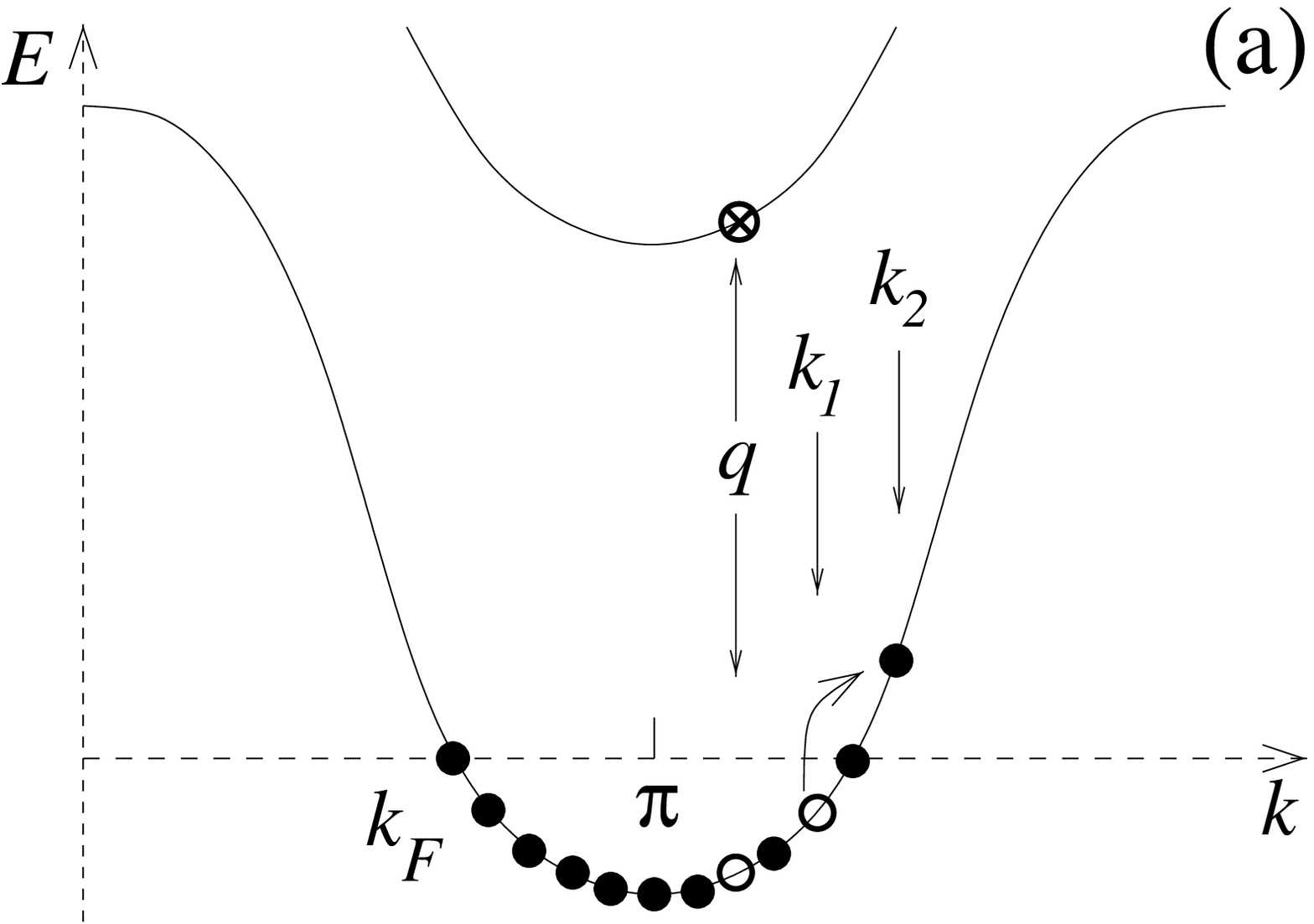}\hspace*{10mm}
\epsfxsize=60mm\epsfbox{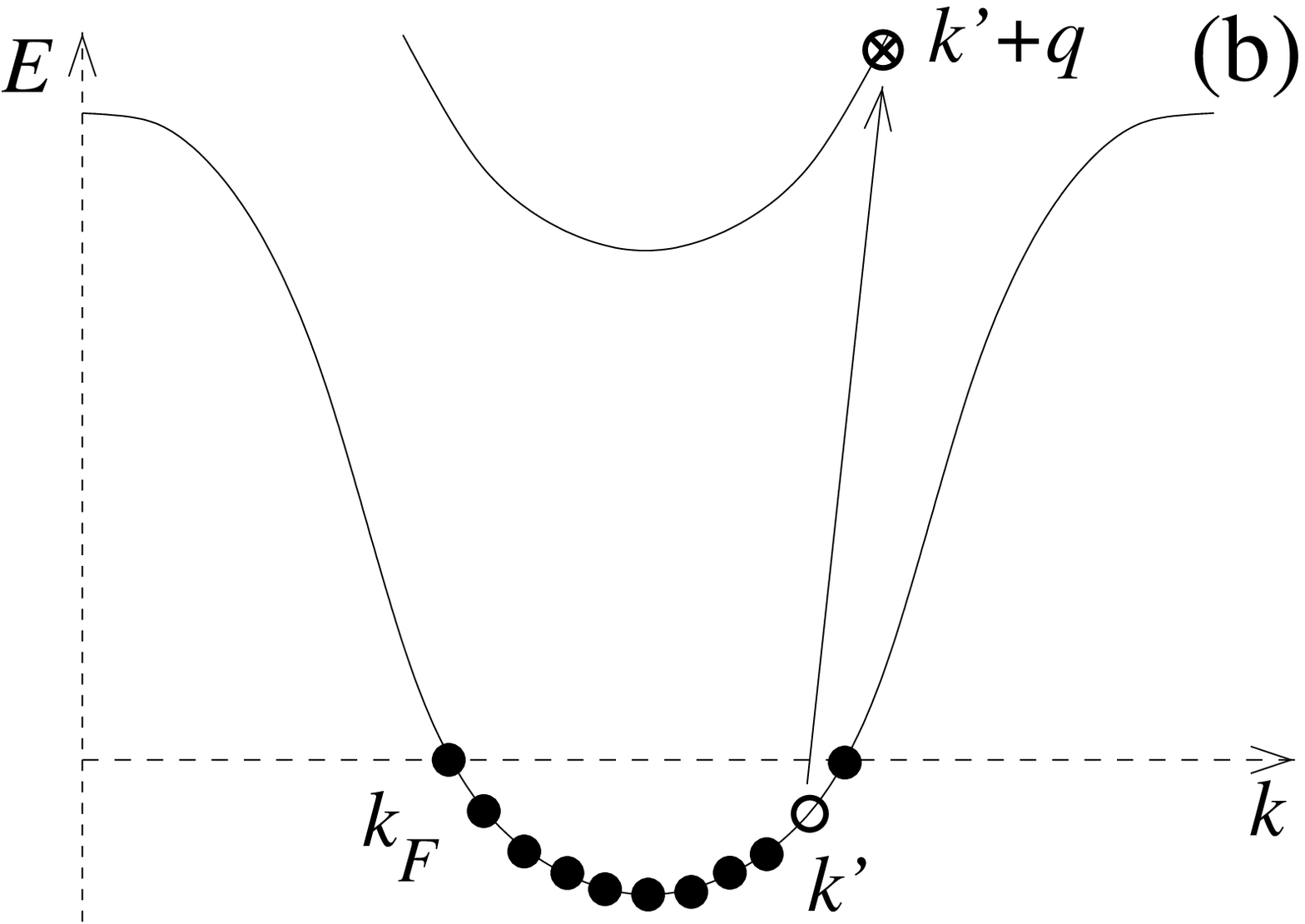}}
\caption{\label{fig:fsea} 
(a) The final-state configuration corresponding
to (B)-type
processes where a new higher-energy boson is added and thus the
allowed wave vectors change because
the total number of particles is changed by one; (b)
schematic view of the ``inter-band'' transitions of (C) type which do
not change the total number of particles.
}
\end{figure}

The matrix element $\langle \mu,\lambda| b^{\dag}_{\mu}(q)
|g.s.\rangle$, $b^{\dag}_{\mu}(q)=L^{-1/2}\sum_{n}b^{\dag}_{n,\mu}
e^{iqn}$,
which determines the contribution to the
response from the (B)-type processes, is nonzero only if the 
selection rules $\lambda=q$, $P'=P+q$ are satisfied, and is
proportional to the determinant of the overlap  matrix:
\begin{equation} 
\label{select}
\langle (\mu,\lambda)_{k_{1}'\cdots k_{N}'}| b^{\dag}_{\mu}(q)
|g.s.\rangle=\delta_{\lambda,q}\delta_{P',P+q}M_{fi},\quad 
M_{fi}=\det\{ \langle \varphi_{i}| \psi_{j}\rangle\}
\end{equation}
Due to the orthogonality catastrophe (OC), the overlap determinant is
generally algebraically vanishing in the thermodynamic limit,
$|M_{fi}|^{2}\propto L^{-\beta}$.  The response is, however, nonzero
and even singular
because there is a macroscopic number of ``shake-up'' configurations
with nearly the same energy.

The value of the OC exponent
$\beta$ is connected to another exponent $\alpha=1-\beta$ which
determines the character of the singularity in the response,
\begin{equation} 
\label{fes} 
I^{B}(q,\omega)=\langle b_{\mu}^{\vphantom{\dag}}
(-q)b_{\mu}^{\dag}(q)\rangle_{\omega}
\propto 1/(\omega-\omega_{\mu}(q))^{\alpha},
\end{equation}
where $\omega_{\mu}(q)$
is the minimum energy difference between the ground state and the
excited configuration.

\begin{wrapfigure}{r}{66mm}
\epsfxsize=55mm
\centerline{\epsfbox{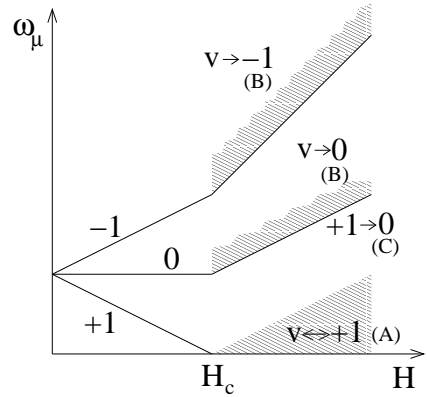}}
\caption{\label{fig:modes} 
The schematic dependence of ``resonance'' lines (peaks in the dynamic structure
factor, shown as solid lines) on the magnetic field. The dashed areas
represent continua. The processes responsible for the 
transitions are indicated near the corresponding lines, e.g. $\rm
v\to-1$ denotes the (B)-type process of creating one boson with
$S^{z}=-1$ from the vacuum, etc.
}
\end{wrapfigure}

For $q=\pi$ the lowest energy excited configuration is symmetric
against $k=\pi$ and is given by $\lambda=\pi$, $k'_{j}=\pi\pm
{2\pi\over L}j$, $j=1,\ldots, N/2$. If $q$ deviates from $\pi$,
$\lambda$ must follow $q$, and in order to satisfy the selection rules
one has to create an additional particle-hole pair  to compensate
the unwanted change of momentum (see Fig.\ \ref{fig:fsea}a). It is easy
to see that for $q$ slightly larger than $\pi$ the lowest energy 
is achieved when $k_{2}=2\pi-k_{F}$
and $k_{1}=2\pi-k_{F}-(q-\pi)$; respectively, for $q$ slightly smaller
than $\pi$one
has $k_{2}=k_{F}$ and $k_{1}=k_{F}+(\pi-q)$. 
As $q$ moves further from
$\pi$, this configuration does not necessarily have the
lowest energy. Indeed,
as we have discussed in Ref.\ \cite{KM02}, there are also other
possible configurations with generally large number of umklapp-type of
particle-hole pairs, whose energy may be lower,
but their contribution to the response can be 
neglected because the corresponding OC exponent is larger than $1$ for this
type of configurations.

Thus, the configuration described above (and its ``shake-up''
perturbations) will give the main contribution to the response for 
$k_{F}<q<2\pi-k_{F}$. The energy difference $\omega(q)$ with the ground state
configuration which determines the position of singularity in
(\ref{fes}) is given by
\begin{equation} 
\label{sing-in-sea} 
\omega_{\mu}(q)=\varepsilon_{\mu}+4t\cos k_{F} -2t\cos(k_{F}+|q-\pi|),\quad
q\in [k_{F},2\pi-k_{F}]\,.
\end{equation}
For $q$ outside the Fermi sea the
symmetric configuration with the hole at $k=\pi$ gives the main
contribution,\cite{KM02} and the corresponding energy difference is
\begin{equation} 
\label{sing-off-sea} 
\omega_{\mu}(q)=\varepsilon_{\mu}+2t\cos q +2t(1+\cos k_{F}),\quad
q\notin [k_{F},2\pi-k_{F}]\,.
\end{equation}
The OC exponent $\beta$ is in both cases the same and equal to $1/2$,
so that the singularity is of the square-root type. 

{\em The quantity $\omega_{\mu}(q)$, which determines the position of
the peak in the response and is normally interpreted as the energy of
the corresponding mode with $S^{z}=\mu$, has a counter-intuitive
dependence on the magnetic field:} indeed, since the definition of
$k_{F}$ is $\varepsilon_{+1}+2t\cos k_{F}=0$, it is easy to see that
e.g. for $q=\pi$ 
{\em one has
$\omega_{\mu}(q)=\varepsilon_{\mu}-\varepsilon_{+1}\propto (1-\mu)H$
instead of $-\mu H$ as one would expect.}  
The resulting picture of
modes which should be seen e.g. in the INS experiment is schematically
shown in Fig.\ \ref{fig:modes}.  This effect may explain the ESR lines
with the resonance energy proportional to $2H$, which were observed in
the high-field phase of NDMAP \cite{Honda+unpub} (such lines would be
normally interpreted as ``forbidden'' transitions with $\Delta
S^{z}=2$).  In recent INS experiments on the same
compound,\cite{Zheludev+02} the change of the slope in the field
dependence of the $q=\pi$ gap for the $S^{z}=0$ mode from zero for
$H<H_{c}$ to $1$ for $H>H_{c}$ was observed, which is also consistent
with the above picture. One should have in mind, however, that this
change of the slope from $-\mu$ to $1-\mu$ is obtained in the model of
noninteracting hardcore bosons, and it will be renormalized by
eventually present interactions.

At finite temperatures the edge singularity becomes damped. In the
vicinity of the singularity, i.e. for $\omega\to\omega_{\mu}(q)$, one
can deduce qualitatively the behavior of $S^{B}(q,\omega)$ from the
formula derived by Ohtaka and Tanabe \cite{OhtakaTanabe90} for the
edge singularity in the photoemission spectrum in case of the contact
core hole potential
\begin{equation} 
\label{B-T} 
I^{B}(q,\omega) ={C(q)\over 4\pi T}
 {(1-\beta)e^{-\beta\gamma(\beta)}\over \Gamma(\beta)}
\left[{2\pi T (D+\bar{D})\over
D\bar{D}}\right]^{\beta}
e^{\widetilde{\omega}/2T}
\left|\Gamma\Big({\beta\over2}
+i{\widetilde{\omega}\over2\pi T}\Big)\right|^{2}, 
\end{equation}
where $\beta=1/2$ is the OC exponent, $\widetilde{\omega}\equiv
\omega-\omega_{\mu}(q)$, 
$\bar{D}=\varepsilon_{+1}-2t$ has the sense of  the energetic depth
of the  Fermi sea, $D=\varepsilon_{+1}+2t$ is the width of the rest of
the $b_{+1}$ band, $\gamma(\beta)$ is defined as 
$\gamma(\beta)=\gamma+\sum_{n=2}^{\infty}{\zeta(2n-1)-1\over n}\beta^{k-1}$,
$\gamma$ being the Euler constant and $\zeta(z)$ the Riemann zeta function,
and, finally, $C(q)$ is the 
 $q$-dependent factor which takes into account the modification
of the overlap determinant $M_{fi}$ in (\ref{select}) due to the
presence of ``compensating'' particle-hole excitations in the final
state for $q\in [k_{F},2\pi-k_{F}]$ (it is identically $1$ for
$q\notin [k_{F},2\pi-k_{F}]$):
\begin{eqnarray} 
\label{factor1} 
&& C(q)=F^{2}(\lambda_{1}\lambda_{2},\rho_{1}\rho_{2}),\quad
F(\lambda_{1}\lambda_{2},\rho_{1}\rho_{2})=F(\lambda_{1}\rho_{1})
F(\lambda_{2}\rho_{2})
-F(\lambda_{1}\rho_{2})F(\lambda_{2}\rho_{1})\nonumber\\
&& \lambda_{1}=\widetilde{k_{F}}+\pi-q,\quad 
\lambda_{2}=q,\quad 
\rho_{1}=\widetilde{k_{F}},\quad \rho_{2}=\pi,\\
&& F(\lambda \rho)={\pi \over L(\lambda-\rho)}
\prod_{m=1}^{N[\lambda]} \Big(1+{1\over2m}\Big)
\prod_{m=1}^{N[\rho]} \Big(1-{1\over2m}\Big)\,,\nonumber
\end{eqnarray}
where $\widetilde{k_{F}}=k_{F}$ if $q<\pi$ and $2\pi-k_{F}$ if $q>\pi$,
and $N[x]$ is defined as $N[x]=L|\widetilde{k_{F}}-x|/2\pi$.
Here $\lambda$ and $\rho$ denote the discrete momenta of ``holes'' and
``particles'', and for $\rho\to\lambda$ one has $F(\lambda\rho)\to1$. 
One can see that for $q\to\pi$, as well as for $q\to k_{F}$ or $q\to
2\pi-k_{F}$, the factor $C(q)\to1$,
and it decays very rapidly if $q$ moves away from those points inside
the $[k_{F},2\pi-k_{F}]$ interval.

\subsection{Transitions of the (C) type}

So far we have been considering the (B)-type transitions. The
dynamical structure factor for $H>H_{c}$ will also have a contribution
from (C)-type transitions corresponding to the transformation of
$b_{+1}$ bosons into $b_{0}$ ones. Those processes do not change the
total number of particles and thus do not disturb the allowed values
of the wave vector, so that there is no orthogonality catastrophe in
this case.  The action of the total spin operator $S^{-}(q)$ is in
this case proportional to that of
$R(q)={1\over\sqrt{L}}\sum_{n}b^{\dag}_{0}b^{\vphantom{\dag}}_{+1}e^{iqn}$. It
is easy to see that the action of $R(q)$ on a state (\ref{eig})
characterized by the set of wave numbers $\{k_{1}\cdots k_{N}\}$ is
given by
\begin{equation} 
\label{Smact}
R(q)|\Psi_{k_{1}\cdots k_{N}}\rangle =
\sqrt{N\over L}\sum_{x_{0}}
{1\over\sqrt{L}}e^{i(P+q)x_{0}}
 \left|\sum_{m=1}^{N}(-)^{m+1} 
|x_{0}\rangle_{ k_{1}\cdots k_{m-1}k_{m+1}\cdots k_{N}}\right|,
\end{equation}
where $x_{0}$ denotes the position of the created $b_{0}$ particle and
the rest of notations is as in (\ref{exc}). The corresponding matrix
element has a very simple form,
\begin{eqnarray} 
\label{matel} 
\langle (\mu,\lambda)_{k_{1}'\cdots k_{N-1}'}|
R(q)|\Psi\{k_{1}\cdots k_{N}\}\rangle
=\sqrt{N/L} \,\delta_{P',P+q}
\delta_{\lambda,k_{m}+q},
\end{eqnarray}
and the problem of calculating the response is thus equivalent to
that for the 1D Fermi gas, with the only difference that we have to
take into account the additional change in energy
$\varepsilon_{0}-\varepsilon_{+1}$ which takes place in the transition
(see Fig.\ \ref{fig:fsea}b).
One can use the well-known formula for
the susceptibility\cite{Lovesey} 
to obtain the  contribution $I^{C}(q,\omega)=\langle R^{\dag}(-q)
R(q)\rangle_{\omega}$  of (C) type processes
into the response:
\begin{equation} 
\label{typeC}
I^{C}(q,\omega)={1\over 1-e^{-\omega/T}}{k_{F}\over 2\pi^{2}}\int
dk\,\big\{ n[\varepsilon_{+1}(k)]-n[\varepsilon_{0}(k+q)]\big\}
\delta(\omega-\varepsilon_{0}(k+q)+\varepsilon_{+1}(k) ),
\end{equation}
where $\varepsilon_{\mu}(k)=\varepsilon_{\mu}+2t\cos k$, and
$n(\varepsilon)=(e^{\varepsilon/T}+1)^{-1}$ is the Fermi distribution
function. 
This contribution contains a square-root singularity, which survives
even for finite temperature and
is located at 
\begin{equation} 
\label{C-sing}
\omega=\varepsilon_{0}-\varepsilon_{+1}+2t\sqrt{2(1-\cos q)} .
\end{equation}
It is interesting to note that though $I^{C}(q,\omega)$
generally does not contain
any quasiparticle contribution,
there is an exception: it transforms into  a $\delta$-function as
$q\to0$. 

\section{Summary}

The dynamical structure factor $S(q,\omega)$ of a gapped
one-dimensional spin system in the high-field critical phase is
studied in the hardcore boson approximation. It is shown that the
presence of a ``condensate'' (Fermi sea) of $S^{z}=+1$ 
triplets in the ground state in the
high-field phase leads to interesting peculiarities in the
contributions to $S(q,\omega)$ from the high-energy excitations
(uncondensed triplet components with $S^{z}=0,-1$).  
The energy of high-energy triplets
gets renormalized because of the presence of the  Fermi sea, which
leads to a substantial change in the magnetic field dependence of the
triplet energies, typically from $-HS^{z}$ to $H(1-S^{z})$. 
Extending our previous study
\cite{KM02}, we show that such peculiarities in the response persist for
any $q$, and can therefore be observed in INS experiments.

{\em Acknowledgments.---}
This work
was partly supported by Volkswagen-Stiftung through Grant No. I/75895.


\begin{thebibliography}{99}

\bibitem{SakaiTakahashi91} T. Sakai and M. Takahashi,
 Phys. Rev. B {\bf 43} (1991), 13383. 

\bibitem{Chaboussant+98a} G. Chaboussant, Y. Fagot-Revurat,
M.-H. Julien, M. E. Hanson, C. Berthier, M. Horvati{\'c},
L. P. L{\'e}vy, and O. Piovesana, Phys. Rev. Lett. {\bf 80} (1998), 2713.

\bibitem{Chaboussant+98} G. Chaboussant, M.-H. Julien,
Y. Fagot-Revurat, M. Hanson, L. P. L{\'e}vy, C. Berthier, M. Horvati{\'c}, and
O. Piovesana, Eur. Phys. J. B {\bf 6} (1998), 167.

\bibitem{Honda+98} Z. Honda, H. Asakawa, and K. Katsumata,
Phys. Rev. Lett. {\bf 81} (1998), 2566.

\bibitem{Honda+99} Z. Honda, K. Katsumata,
M. Hagiwara, and M. Tokunaga, Phys. Rev. B {\bf 60} (1999), 9272.

\bibitem{Orendac+99} M. Orend{\'a}\v{c}, S. Zvyagin, 
A. Orend{\'a}{\v{c}}ov{\'a}, M. Seiling, B. L{\"u}thi, A. Feher and
M. W. Meisel, Phys. Rev. B {\bf 60} (1999), 4170. 


\bibitem{Totsuka98} K. Totsuka, Phys. Rev. B {\bf 57} (1998), 3454.

\bibitem{Mila98} F. Mila, Eur. Phys. J B {\bf 6} (1998), 201.



\bibitem{ChitraGiamarchi97} R. Chitra and T. Giamarchi, Phys. Rev. B
{\bf 55} (1997), 5816.

\bibitem{GiamarchiTsvelik99}  T. Giamarchi and A. M. Tsvelik,
Phys. Rev. B {\bf 59}, 11398 (1999).

\bibitem{FurusakiZhang99} A. Furusaki and S.-C. Zhang, Phys. Rev. B
{\bf 60}, 1175 (1999).

\bibitem{HikiharaFurusaki01} T. Hikihara and A. Furusaki, Phys. Rev. B
{\bf 63} (2001), 134438.

\bibitem{Bocquet+01} M. Bocquet, F. H. L. Essler, A. M. Tsvelik, and
A. O. Gogolin, Phys. Rev. B {\bf 64} (2001), 094425.

\bibitem{KonikFendley01} R. Konik and P. Fendley, preprint
cond-mat/0106037. 


\bibitem{KMS99} A. K. Kolezhuk, H.-J. Mikeska, and U. Schollw\"ock,
Phys. Rev. B {\bf 59} (1999), 13565.

\bibitem{Honda+unpub} Z. Honda, private communication.

\bibitem{Zheludev+02} A. Zheludev, Z. Honda, Y. Chen, C. L. Broholm,
K. Katsumata, and S. M. Shapiro, preprint cond-mat/0107416.

\bibitem{Ruegg+02} Ch. R\"uegg, N. Cavadini, A. Furrer, K. Kr\"amer,
H. U. G\"udel, P. Vorderwisch, and H. Mutka, preprint (to appear in
Appl. Phys. A).

\bibitem{AKLT} I. Affleck, T. Kennedy, E. H. Lieb and H. Tasaki, Phys.
Rev. Lett. {\bf 59} (1987), 799.\\
Commun. Math. Phys. {\bf 115} (1988), 477.

\bibitem{KM02} A. K. Kolezhuk and H.-J. Mikeska, Phys. Rev. B {\bf 65}
(2002), 014413.

\bibitem{FathSolyom93} G. F\'ath and J. S\'olyom, J. Phys.:
Condens. Matter {\bf 5} (1993), 8983.\\
 N. Elstner and H.-J. Mikeska, Phys. Rev. B {\bf 50} (1994), 3907.\\
U. Neugebauer and H.-J. Mikeska, Z. Phys. B {\bf 99} (1996), 151.

\bibitem{Fannes+89} M. Fannes, B. Nachtergaele and R. F. Werner,
Europhys. Lett. {\bf 10} (1989), 633.\\
 Commun. Math. Phys. {\bf 144} (1992), 443.

\bibitem{Klumper+91-93} A. Kl\"umper, A. Schadschneider and
J. Zittartz, J. Phys.  A {\bf 24} (1991), L955.\\
 Z. Phys. B {\bf 87} (1992), 281.\\
 Europhys. Lett. {\bf 24} (1993), 293.

\bibitem{TotsukaSuzuki95} K. Totsuka and M. Suzuki, J. Phys.:
Condens. Matter {\bf 7} (1995), 1639.

\bibitem{Krohn} M. Krohn, Diploma thesis, University of Hannover (2000). 

\bibitem{Schulz+83-86} H. J. Schulz and C. Bourbonnais, Phys. Rev. B
{\bf 27} (1983), 5856.\\
H. J. Schulz, Phys. Rev. B {\bf 34} (1986), 6372.

\bibitem{Barzykin01} V. Barzykin, Phys. Rev. B {\bf 63} (2001), 140412(R).

\bibitem{Mueller+81} G. M\"uller, H. Thomas, H. Beck, and
J. C. Bonner, Phys. Rev. B {\bf 24} (1981), 1429.

\bibitem{Derzhko+00} O. Derzhko, T. Krokhmalskii, and J. Stolze,
J. Phys. A: Math. Gen. {\bf 33} (2000), 3063.

\bibitem{CastellaZotos93} H. Castella and X. Zotos, Phys. Rev. B {\bf
47} (1993), 16186.

\bibitem{Anderson67} P. W. Anderson, Phys. Rev. Lett. {\bf 18} (1967), 1049.\\
 Phys. Rev. {\bf 164} (1967), 352.

\bibitem{OhtakaTanabe90} 
K. Ohtaka and Y. Tanabe, Rev. Mod. Phys. {\bf
62} (1990), 929.

\bibitem{Lovesey} See, e.g., S. W. Lovesey, {\em Theory of neutron scattering
from condensed matter. Vol. 1} (Clarendon Press, Oxford 1984) p. 231.



\end{thebibliography}
\end{document}